\pdfoutput=1
\documentclass[aip, jcp, reprint,floatfix,a4paper]{revtex4-1}  
\usepackage{graphicx}
\usepackage{dcolumn}
\usepackage{bm}

\usepackage[utf8]{inputenc}
\usepackage[T1]{fontenc}
\usepackage{mathptmx}
\usepackage{color}
\usepackage{tikz}
\usepackage{amsmath}
\usepackage{geometry}
\usepackage{amssymb}
\usepackage{csquotes}
\usepackage{enumitem}
\usepackage{siunitx}
\usepackage{placeins}

\usepackage{hyperref}
\hypersetup{
        colorlinks=true,
        citecolor=black,
        filecolor=black,
        linkcolor=black,
        urlcolor=black
}

\newcommand{\vecx}{\mathbf{x}}

\newcommand{\vecr}{r}
\newcommand{\br}[1]{\mathopen{}\left(#1\right)\mathclose{}}
\newcommand{\abs}[1]{\left|#1\right|}

\begin{document}

\preprint{arXiv}

\title{Targeted free energy estimation via learned mappings}

\affiliation{ 
    DeepMind, London, United Kingdom
}

\author{Peter Wirnsberger}
\thanks{These two authors contributed equally and are correspondence authors: \{pewi, aybd\}@google.com.}
\author{Andrew J.\ Ballard}
\thanks{These two authors contributed equally and are correspondence authors: \{pewi, aybd\}@google.com.}
\author{George Papamakarios}
\author{Stuart Abercrombie}
\author{Sébastien Racani\`ere}
\author{Alexander Pritzel}
\author{Danilo Jimenez Rezende}
\author{Charles Blundell}

\date{\today}

\begin{abstract}
Free energy perturbation (FEP) was proposed by Zwanzig more than six decades ago as a method to estimate free energy differences, and has since inspired a huge body of related methods that use it as an integral building block. Being an importance sampling based estimator, however, FEP suffers from a severe limitation: the requirement of sufficient overlap between distributions. One strategy to mitigate this problem, called Targeted Free Energy Perturbation, uses a high-dimensional mapping in configuration space to increase overlap of the underlying distributions. Despite its potential, this method has attracted only limited attention due to the formidable challenge of formulating a tractable mapping. Here, we cast Targeted FEP as a machine learning problem in which the mapping is parameterized as a neural network that is optimized so as to increase overlap. We develop a new model architecture that respects permutational and periodic symmetries often encountered in atomistic simulations and test our method on a fully-periodic solvation system.
We demonstrate that our method leads to a substantial variance reduction in free energy estimates when compared against baselines, without requiring any additional data. 
\end{abstract}

\maketitle

\section{\label{sec:introduction}Introduction}
Free energy estimation is of central importance in the natural sciences. Accurate estimation of free energies, however, is challenging, as many systems are out of reach of experimental methods and analytic theory. Computer-based estimation has thus emerged as a valuable alternative. Successful application areas of in-silico free energy estimation span industry and scientific research,
including drug discovery \cite{shirts2010}, condensed matter physics~\cite{Auer2001}, materials science~\cite{Damasceno2012}, structural biology~\cite{Curk2018}, and the effects of mutagenesis~\cite{Hauser2018:CommBio}.
Because of its importance and wide ranging applications, computer-based free energy estimation has been an active field of research for decades~\cite{cchipot07:molsim}.

At the core of many state-of-the-art estimators~\cite{Shirts2008} lies the free energy perturbation (FEP) identity introduced by Zwanzig~\cite{Zwanzig1954} in 1954: 
\begin{equation}
    \mathbb{E}_A\left[ e^{-\beta \Delta U} \right] = e^{-\beta \Delta F}.
\label{eq:fep}
\end{equation}
Here $\Delta F = F_B - F_A$ is the Helmholtz free energy difference between two thermodynamic states $A$ and $B$, each connected to a thermal reservoir at inverse temperature $\beta$. We denote $\vecx$ as a point in the system's configuration space, and define 
\begin{equation}
    \Delta U(\vecx) = U_B(\vecx) - U_A(\vecx),
    \label{eq:dUa}
\end{equation}
with $U_A = U(\vecx; \lambda_A)$ and $U_B = U(\vecx; \lambda_B)$ the energy functions associated with $A$ and $B$. The parameters $\lambda_A$ and $\lambda_B$ could, for example, represent coupling coefficients of a particle-particle interaction potential.
By $\mathbb{E}_A[\cdots]$ we denote an expectation under equilibrium distribution $\rho_A \propto e^{-\beta U_A}$ (and similarly for $B$). A procedure for computation of $\Delta F$ via Eq.~(\ref{eq:fep}) is then as follows: a number of samples  are first drawn from $\rho_A$, for example via a Markov chain Monte Carlo or molecular dynamics simulation. The change in energy, $\Delta U$, associated with instantaneously switching $\lambda_A \rightarrow \lambda_B$ is then computed, from which an exponential average is taken. From a statistical point of view, FEP is an application of Importance Sampling (IS), where $\rho_A$ serves as the proposal distribution~\cite{Neal2001,arnaud2001sequential,liu2008montecarlo}. 

While Eq.~(\ref{eq:fep}) is exact, the convergence of this estimator for a finite number of samples strongly depends on the degree to which $A$ and $B$ overlap in configuration space~\cite{Pohorille2010:JPCB}. Indeed, the dominant contributions to the above expectation will come from samples of $A$ that are typical under $B$, and such contributions become increasingly rare with decreasing overlap~\cite{jarzynski06:PRE}.

\begin{figure}[htbp]
    \centering
    \includegraphics[width=.95\columnwidth]{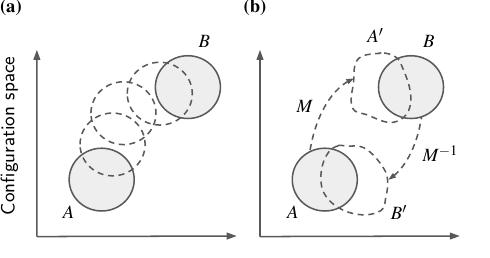}
    \caption{\textbf{The overlap problem.} Efficient estimates of free energy differences rely on a decent configuration-space overlap between equilibrium distributions. Each panel represents a set of distributions on configuration space, with the circles indicating regions of appreciable mass. (a) In the commonly-used multi-staged approach, a chain of intermediate states is defined between $A$ and $B$ such that a decent overlap exists between each neighboring pair. Free energy differences are computed with respect to each neighboring pair, and summed to obtain the difference of interest. (b) In the targeted approach, an invertible mapping $M$ transports the equilibrium distribution $A$ to a new distribution $A^\prime$. Convergence is improved upon if the mapping results in an increased overlap with $B$ and similarly for the reverse direction.}
\label{fig:overlap}
\end{figure}

There exist multiple strategies for mitigating the overlap requirement. Arguably the most common strategy is a multi-staged approach, also known as stratification, in which a sequence of intermediate thermodynamic states is defined between $A$ and $B$ (Fig.~\ref{fig:overlap}a). Here the increased convergence is facilitated by demanding that neighboring pairs of states be chosen to contain sufficient overlap. The quantity of interest, $\Delta F$, is then recovered as a sum over the pairwise differences $\Delta F_{i, i+1}$~\cite{cchipot07:molsim}. The Multistage Bennett Acceptance Ratio~\cite{Shirts2008} (MBAR) estimator is a prominent example of an estimator that follows this strategy. However, multi-staged approaches require samples from multiple states as well as a suitable order parameter to define intermediate stages. Furthermore, it is unclear a priori how best to discretize the order parameter or how many stages to use.

An alternative, elegant strategy to increasing overlap is by incorporating configuration space maps. Jarzynski developed Targeted Free Energy Perturbation~\cite{jarzynski02:pre} (TFEP), a generalization of FEP whereby an invertible mapping defined on configuration space transports points sampled from $A$ to a new distribution, $A^{\prime}$ (see Fig.~\ref{fig:overlap}b). Jarzynski showed that a generalized FEP identity can be applied to this process, from which the free energy difference can be 
recovered. Importantly, if the mapping is chosen wisely, an effective overlap can be increased, leading to quicker convergence of the TFEP estimator. Hahn and Then extended TFEP to the bidirectional setting~\cite{Hahn09:PRE}, whereby the mapping and its inverse are applied to samples from $A$ and $B$, respectively. Lower-error free energy estimates can then be obtained via the statistically-optimal BAR estimator~\cite{Bennett1976}.

Whether in the unidirectional or bidirectional case, the main challenge for targeted approaches is crafting a mapping that is capable of increasing overlap. Unfortunately for most real-world problems the physical intuition needed to develop such a technique is simply lacking. Modern-day machine learning (ML) techniques, however, seem perfectly suited for this task.

Since the introduction of TFEP in 2002, research in ML has made remarkable progress in fields of image classification~\cite{krizhevsky12:NeurIPS,he2016resnet}, playing video~\cite{Mnih2015} and board games~\cite{Silver2016:Nature, Silver2018:Science}, and generative modelling of images~\cite{brock2018large,karras2019stylegan}. ML has also enabled advances in the natural sciences, including state of the art protein structure prediction~\cite{Senior2020}, neural-network based molecular force fields~\cite{Morawietz2016, Zhang2018}, generative modelling of lattice field theories~\cite{Albergo2019}, new paradigms for sampling equilibrium distributions of molecules~\cite{Noe19:Science}, and variational free energy estimates~\cite{Wu2019, Li2018}.

In this work, we turn targeted free energy estimation into a machine learning problem. In lieu of a hand-crafted mapping, we represent our mapping by a deep neural network whose parameters are optimized so as to maximize overlap. 
Once trained, the free energy can then be computed by evaluating the targeted estimator with our learned mapping. 
Below we will consider both unidirectional and bidirectional settings, and will refer to them as \textit{Learned Free Energy Perturbation} (LFEP) and \textit{Learned Bennett Acceptance Ratio} (LBAR), respectively\@. 

A key contribution of this work is the development of a mapping that respects the underlying symmetries of our system of study. In particular, our neural network is equivariant to permutation of identical particles and respects periodic boundary conditions by construction. These are particularly important considerations when modelling atomic systems, as they often obey such symmetries.

The rest of our manuscript is structured as follows. In Sec.~\ref{sec:theory}\@ below we  summarize the previously-developed targeted free energy estimators that we will be making use of. This is followed by development of suitable training objectives to maximize overlap (Sec.~\ref{sec:loss}). We then demonstrate our method by applying it to a solvation system. In Sec.~\ref{sec:setup}, we describe the experimental setup and discuss inherent symmetries that are exploited to devise a model with the correct inductive biases (Sec.~\ref{sec:model})\@. Finally, we present experimental results in Sec.~\ref{sec:results} and discuss our findings in Sec.~\ref{sec:discussion}.

\section{\label{sec:theory}Theoretical background}
In the following we will refer to $A$ and $B$ as the thermodynamic states, defined by equilibrium densities $\rho_A(\vecx) = e^{-\beta U_A(\vecx)} / Z_A$ and $\rho_B(\vecx) = e^{-\beta U_B(\vecx)}/Z_B$, where $Z_A$ and $Z_B$ are the normalization constants (partition functions). In the targeted scheme, configurations $\mathbf{x}$ are
drawn from $A$ and mapped to new configurations
$\mathbf{y} = M (\mathbf{x})$ via an invertible, user-specified mapping $M$. The set of mapped configurations can be thought of as samples from a new state $A^{\prime}$: 
\begin{equation}
    M : A \rightarrow A^{\prime}.
\end{equation}
Similarly, we also consider the reverse case where 
configurations are drawn from $B$ and mapped to $B^{\prime}$ via the inverse
\begin{equation}
    M^{- 1} : B \rightarrow B^{\prime}. 
\end{equation}
We refer to this pair of prescriptions as the \enquote{forward} and \enquote{reverse}
process, respectively. For each process we denote generalized energy
differences as
\begin{align}
  \Phi_F ( \mathbf{x} ) & = U_B ( M ( \mathbf{x}
  ) ) - U_A ( \mathbf{x} ) - \beta^{- 1} \log \abs{J_M
  ( \mathbf{x} )},   \label{eq:phi_f} \\
  \Phi_R ( \mathbf{x} ) & = U_A \br{ M^{- 1} (
  \mathbf{x} )} - U_B ( \mathbf{x}) - \beta^{- 1}
  \log{\abs{J_{M^{- 1}} ( \mathbf{x} )}},
  \label{eq:phi_r}
\end{align}
where $J_M$ and $J_{M^{- 1}} =J_M^{-1}$ are the Jacobian determinants
associated with the mappings. As originally shown by Jarzynski~\cite{jarzynski02:pre}, an identity exists which relates $\Delta F$ to an ensemble of realizations of
$\Phi_F$:
\begin{equation}
    \mathbb{E}_A \left[e^{- \beta \Phi_F}\right] = e^{- \beta \Delta F} . \label{eq:tfep}
\end{equation}
Eq.~(\ref{eq:tfep}) can be regarded as a generalization of FEP, as it holds for any invertible $M$, and reduces to Eq.~(\ref{eq:fep}) if $M$ is the identity. An analogous equation holds for the reverse process. Derivations of Eq.~(\ref{eq:tfep}) can be found in Refs.~{\onlinecite{jarzynski02:pre,Hahn09:PRE}} or Appendix~\ref{app:tfepder}.

Hahn and Then extended the above result to the bidirectional case~\cite{Hahn09:PRE}, showing a
fluctuation theorem (FT) exists between the forward and reverse processes:
\begin{equation}
    \frac{p_F (\phi)}{p_R (- \phi)} = e^{\beta (\phi - \Delta F)}.
    \label{eq:ft}
\end{equation}
The functions
\begin{equation}
    p_F (\phi) = \int \mathrm{d} \mathbf{x}\ \rho_A ( \mathbf{x} )
   \delta \br{ \phi - \Phi_F ( \mathbf{x} ) }
   \label{eq:pF}
\end{equation}   
and
\begin{equation}
    p_R (\phi) = \int \mathrm{d} \mathbf{x}\ \rho_B ( \mathbf{x} )
   \delta \br{ \phi - \Phi_R ( \mathbf{x} ) }
   \label{eq:pR}
\end{equation}
can be thought of as generalized work distributions associated with the
mapping processes and $\delta$ is the Dirac delta function. With these bidirectional estimates, Bennett's Acceptance Ratio (BAR) method~\cite{Bennett1976}
can be employed as an alternative estimator of ${\Delta F}$~\cite{Hahn09:PRE}. BAR estimation of $\Delta F$ can be formulated as a self-consistent iteration of the equation
\begin{equation}
    \mathbb{E}_A\left[f(\beta (\Phi_F-\Delta F)) \right] = \mathbb{E}_B\left[f(\beta (\Phi_R+\Delta F))\right],
    \label{eq:bar}
\end{equation}
where $f(x)=1/(1+e^x)$ is the Fermi function. For simplicity in Eq.~(\ref{eq:bar}) we restrict ourselves to the case where the number of samples in the forward and reverse directions are equal, but more general formulations exist. 
BAR has a statistical advantage over FEP as it has been shown to be the minimum variance free energy estimator for any asymptotically-unbiased method~\cite{shirts2003}. Because of this property, BAR is generally the
method of choice when samples from both $A$ and $B$ are available. 

In summary, our method proceeds in two stages by first computing optimized work values using Eqs.~(\ref{eq:phi_f})--(\ref{eq:phi_r}) and then estimating $\Delta F$. For the latter we can employ the generalized FEP estimator (\ref{eq:tfep}) in the unidirectional setting (LFEP), or solve the BAR equations~(\ref{eq:bar}) in the bidirectional setting (LBAR).
This highlights an important difference between LBAR and other maximum likelihood free energy estimators which assume the work values to be fixed~\cite{Bennett1976, Maragakis2006, Shirts2008}. Instead, LBAR learns to optimize the work values and subsequently combines them optimally to predict $\Delta F$.

Crucially, the targeted estimators above hold for \textit{every}
invertible mapping. That is, given an infinite number of samples, any
invertible choice of $M$ will produce a consistent estimate of $\Delta F$. Of
course, the finite-sample convergence properties are of more practical
importance and will strongly depend on the choice of $M$.

\section{\label{sec:loss}Training objective}
In a distributional sense, the forward and reverse processes act to transform $\rho_A$ and $\rho_B$ into $\rho_{A^{\prime}}$ and
$\rho_{B^{\prime}}$, as depicted in Fig~\ref{fig:overlap}. In what follows we will refer
to the distribution of mapped configurations as the ``images'' (i.e.\ $\rho_{A^{\prime}}$\
and $\rho_{B^{\prime}}$), and the distributions we want them mapped towards as the
``targets'' (i.e.\
$\rho_A$\ ($\rho_B$) for the forward (reverse) process). Due to the deterministic mapping, the bases and images are
related by the change of variable formula,
\begin{align}
   \rho_{A^{\prime}} (M (\mathbf{x})) &= \rho_A (\mathbf{x}) / \abs{J_M (\mathbf{x})}   \label{eq:change_of_variableA}
, \\
   \rho_{B^{\prime}} \br{ M^{- 1} ( \mathbf{x} )} &=
   \rho_B (\mathbf{x}) / \abs{J_{M^{- 1}} (\mathbf{x})}   \label{eq:change_of_variableB}
. 
\end{align}

The crucial consideration for
convergence of our estimators (Eq.~\eqref{eq:tfep} and Eq.~\eqref{eq:bar}) is the overlap between the image and target
distributions~\cite{jarzynski02:pre}. Indeed, in the limit that images and targets
coincide, Jarzynski showed~\cite{jarzynski02:pre} that $p_F(\phi) \rightarrow \delta\br{\phi - \Delta F}$. This implies that the convergence of Eq.~(\ref{eq:tfep}) is immediate (i.e. only one sample is needed). 
In this limit, it is also the case that $p_R(\phi) \rightarrow \delta\br{\phi + \Delta F}$ implying that the expectation values on either side of Eq.~\eqref{eq:bar} converge immediately.
This overlap argument is
further reinforced in the Appendix~\ref{app:tfepder}, where we show that the TFEP estimator can
be interpreted as a FEP estimator between $A'$ and $B$.

We now turn our attention to the construction of a loss function that
accurately judges the quality of $M$. Guided by the considerations of overlap,
we consider the Kullback--Leibler~(KL) divergence between the image and target. For the forward
process we have:
\begingroup
\allowdisplaybreaks
\begin{align}
  D_{\text{KL}}[\rho_{A^{\prime}} | | \rho_B] &= \mathbb{E}_{\mathbf{x} \sim
  A^{\prime}} \left[ \log \frac{\rho_{A^{\prime}} ( \mathbf{x} )}{\rho_B
  ( \mathbf{x} )} \right]\nonumber\\
  &= \mathbb{E}_{\mathbf{x} \sim A^{}} \left[ \log \frac{\rho_{A^{\prime}}
  \br{ M ( \mathbf{x} ) }}{\rho_B \br{ M (
  \mathbf{x} ) }} \right]\nonumber\\
  &= \mathbb{E}_{\mathbf{x} \sim A^{}} \left[ \log \frac{\rho_A (
  \mathbf{x} ) / \abs{J_M ( \mathbf{x} )}}{\rho_B \br{ M (
  \mathbf{x} ) }} \right]\nonumber\\
  &= \mathbb{E}_{\mathbf{x} \sim A^{}} \left[ \beta \Phi_F(\mathbf{x}) + \log{ \frac{Z_B}{Z_A}} \right] \nonumber\\
  &= \beta \left(\mathbb{E}_A [\Phi_F] - \Delta F \right).
  \label{eq:lossder}
\end{align}
\endgroup
In the above derivation, we invoked a change of variable formula in going from the first to third lines, and
used the identity $-\beta \Delta F  =\log{Z_B} - \log{Z_A}$ to get to the last. 
An analogous equation can be derived for the reverse process yielding
\begin{equation}
  D_{\text{KL}} [\rho_{B^{\prime}} | | \rho_A] = \beta \left(\mathbb{E}_B [\Phi_R] + \Delta F \right).
  \label{eq:loss_rev}
\end{equation}
While from Eqs.~(\ref{eq:lossder}--\ref{eq:loss_rev}) it is clear
that the KL cannot be accurately estimated unless ${\Delta}F$ is known, in terms
of optimizing, ${\Delta}F$ and $\beta$ can be disregarded as they are
constants.

Below we consider two separate training regimes for our model. In the unidirectional case, the model was trained only on the forward process, with a loss
function  
\begin{equation}
    \label{eq:loss_lfep}
    \mathcal L_{\text{LFEP}} = \mathbb{E}_{A} [\Phi_F].
\end{equation} 
In the bidirectional case, the model was trained using both forward and reverse processes with the loss
\begin{equation}
    \label{eq:loss_lbar}
    \mathcal L_{\text{LBAR}} = \mathbb{E}_{A} [\Phi_F] + \mathbb{E}_{B} [\Phi_R].
\end{equation}
When samples from both states are available, bidirectional training is preferable. Unlike the unidirectional loss, $\mathcal L_{\text{LBAR}}$ explicitly encourages both $\rho_{A^{\prime}}$ and $\rho_{B^{\prime}}$ to be mass-covering~\citep{minka2005divergence}, which is important for good performance of importance-sampling estimators~\cite{neal2005bridgesampling}. 

\section{\label{sec:setup}Experimental setup}
\begin{figure}[htbp]
    \centering
    \includegraphics[width=.7\columnwidth]{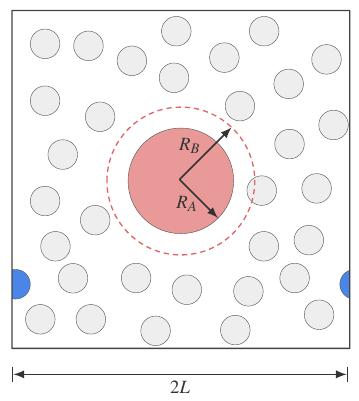}
\caption{\textbf{Illustration of the simulation box.} A solute particle (pink), located at the center, is grown from radius $R_A$ to radius $R_B$ thereby compressing the space accessible to the $N=125$ solvent particles (grey). Opposite faces of the cubic simulation box with edge length $2L$ are associated with each other such that a particle gets inserted again on the opposite side as it tries to leave the reference box (blue). The reference box is therefore topologically equivalent to a torus.}
\label{fig:system}
\end{figure}

To test our method, we consider a system similar to the one used by Jarzynski~\cite{jarzynski02:pre} consisting of a repulsive solute immersed in a bath of $N=125$ identical solvent particles. The task is to calculate the free energy change associated with growing the solute radius from $R_A$ to radius $R_B$ (see Fig.~\ref{fig:system}). In contrast to a hard-sphere solute as used in Ref.~\onlinecite{jarzynski02:pre}, we modeled our solute as a soft sphere. This is because any finite particle overlap would lead to infinite forces, which our training method cannot handle.

Intuitively, an effective mapping should push solvent particles away from the center to avoid high-energy steric clashes with the expanding solute. Jarzynski followed this intuition, defining a mapping that uniformly compresses the solvent particles amidst an expanding repulsive solute. Although this mapping gave a significant convergence gains when applied to a hard solute~\cite{jarzynski02:pre}, it is not directly applicable to soft solutes. This is because the phase space compression results in a transformed density whose support is not equal to that of the target density, violating the assumption of invertibility. 

Below we demonstrate that we no longer need to rely on physical intuition to hand-craft a tractable mapping; this process can be fully automated using the general framework proposed in this work. In order to learn an effective mapping, however, it is crucial that the model be compatible with the inherent symmetries of the underlying physical system.

\subsection{\label{subsec:energy}Energy}
Periodic boundary conditions (PBCs) confine the $N$-particle system to a $3N$-dimensional torus, $\vecx= (\vecr_1, \ldots, \vecr_N) \in \mathbb{T}^{3N}$, where each coordinate of the position vector $\vecr_i \in \mathbb{T}^3$ of particle $i$ lives in the interval $[-L,L]$ (see Fig.~\ref{fig:system}). The total energy can then be decomposed into a sum of pairwise contributions according to
\begin{equation}
U_\alpha(\vecx) = \sum_{i=1}^{N}\sum_{j < i} u(|\vecr_{ij}'|) +\sum_{i=1}^{N} v(|\vecr_{i}|; R_\alpha),
\label{eq:energy}
\end{equation}
where subscript $\alpha \in \{A,B\}$ denotes the state and $\vecr_{ij} = \vecr_j - \vecr_i$. The quantity $\vecr_{ij}' = \vecr_{ij} - 2L\ \text{round}(\vecr_{ij}/(2L))$ is a difference vector whose components correspond to the shortest, signed distance in the respective dimension, and the function round is applied element-wise giving the nearest integral number. The radially symmetric functions $u(r)$ and $v(r)$ represent the Lennard-Jones~\cite{Jones1924} (LJ) and Weeks--Chandler--Andersen~\cite{Weeks1971} (WCA) potentials. In practice, we truncate LJ interactions using a radial cutoff of $L$ and shift the potential to be zero at the cutoff~\cite{Frenkel2002}.

\subsection{\label{subsec:data}Training data}
The data used for training and evaluation was generated via molecular dynamics (MD) simulations using the package LAMMPS~\cite{Plimpton1995}\@. Simulation frames were generated via Langevin dynamics and were saved infrequently enough to ensure decorrelated samples (as judged by the potential energy decorrelation time), giving a total of $\num{5e4}$ frames per simulation. A total of $20$ independent simulation trajectories were generated for each thermodynamic state, starting from randomly-initialized configurations and momenta. Ten independent training and evaluation datasets were then constructed from these simulations by first concatenating and shuffling configurations, and then partitioning them into $N_{\mathrm{train}} = \num{9e4}$ training and $N_{\mathrm{test}} = \num{1e4}$ test samples for each dataset. The value of $N_{\mathrm{train}}$ was chosen such that the baseline BAR estimator, when evaluated on $N_{\mathrm{train}}$ data points, yields a reasonably accurate free energy difference. Below we train and evaluate our model on these datasets so as to compute statistical convergence properties of the estimators across independent runs.  Further simulation details are summarized in Appendix~\ref{app:system}.

\subsection{\label{subsec:symmetries}Symmetries}
The system under consideration exhibits properties that are widely encountered in the atomistic simulation community: periodic boundary conditions (PBCs) and permutation invariance\@. PBCs are usually employed to reduce finite-size effects. Permutation invariance arises as a consequence of the energy being invariant to particle permutations---a condition that is satisfied by the solvent particles as they are all identical.

PBCs are a choice of geometry for the space in which particles live: a 3D torus. Without them, the system (including the solute) would admit rigid rotation, reflection and rigid translation symmetries. With them, only the translation symmetries remain, and a discrete set of rotations/reflections.
The original rigid translations in $\mathbb{R}^3$ become translations by elements of the torus $\mathbb{T}^3$, leaving the energy invariant. Because of this symmetry, we can fix the solute at the origin of the simulation box without affecting the ratio $Z_B/Z_A$ (as shown in Fig.~\ref{fig:system}).
The remaining set of discrete rotations/reflections is the group of symmetries of a cube (the \textit{octahedral group}), which contains only $48$ elements.

We design the mapping $M$ to respect PBCs and permutation invariance by construction. This means that the state $A^{\prime}$ obtained by transforming $A$ via $M$ is guaranteed to have the 3D torus geometry stipulated by PBCs, and to be symmetric with respect to any permutation of solvent particles. 
The next section discusses in detail how these symmetries are implemented in the model architecture.

Our model architecture does not obey the octahedral symmetries, in the sense that $A^{\prime}$ is not guaranteed to be symmetric with respect to the $48$ permutations and/or reflections of the three coordinate axes. However, since the size of this symmetry group is small, we account for the octahedral symmetries via \textit{training-data augmentation} instead. That is, during training we transform every training data point (system configuration) by a random element of the octahedral group. As training progresses, all $48$ transformations of each data point are likely to be seen by the model, thus the model is trained to learn these symmetries from data. In comparison, exhausting the $N! = 125!$ permutation symmetries by training-data augmentation would be infeasible in any reasonable training time.

\section{\label{sec:model}Model}
We implement the mapping $M$ using a deep neural network, parameterized by a set of learnable parameters $\theta$. In designing the architecture of the network, we take into account the following considerations.
\begin{enumerate}[label = (\alph*), leftmargin=*]
    \item The mapping $M$ must be bijective, and the inverse mapping $M^{-1}$ should be efficient to compute, for any setting of the parameters $\theta$.
    \item The Jacobian determinant $J_M$ should be efficient to compute for any setting of $\theta$.
    \item The network should be flexible enough to represent complex mappings.
    \item The transformed distributions $\rho_{A'}$ and $\rho_{B'}$ should respect the boundary conditions and symmetries of the physical system.
\end{enumerate}

The first three requirements are satisfied by a class of deep neural networks known as \textit{normalizing flows}~\cite{papamakarios2019normalizing}, which are invertible networks with efficient Jacobian determinants. Since bijectivity is a closed property under function composition, multiple normalizing flows (or ``layers'') can be composed into a deeper flow, yielding a model with increased flexibility. We implement $M$ as a normalizing flow composed of $K$ invertible layers, that is,
\begin{equation}
M=M_K\circ\cdots\circ M_1.     
\end{equation}
Each layer $M_k:\mathbb{T}^{3N}\rightarrow\mathbb{T}^{3N}$ is of the same architecture but it has its own learnable parameters $\theta_k$, and the learnable parameters of $M$ are simply $\theta = (\theta_1, \ldots, \theta_K)$.

\begin{figure}[htbp]
    \centering
    \includegraphics[width=.98\columnwidth]{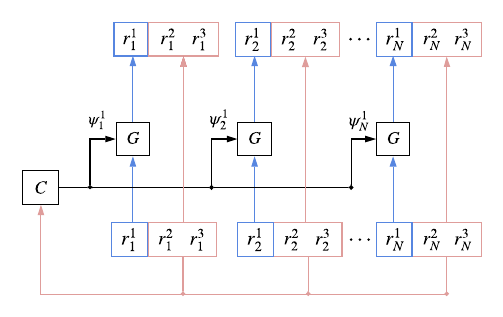}
    \caption{\textbf{Illustration of a coupling layer $M_k$.} 
    A subset of coordinates, here $\mathbb{I}_k = \{2,3\}$, remains unchanged 
    while all other coordinates are transformed by a \textit{circular spline} $G$. 
    The parameters $\psi_i^{\nu}$ of $G$ are produced by a separate model $C$, 
    which we implemented using the \textit{transformer} architecture~\cite{vaswani2017transformer}.}
    \label{fig:coupling_layer}
\end{figure}

Our implementation of $M_k$ is based on the architecture proposed by Dinh et al.~\cite{dinh2017density}, which is often referred to as the \textit{coupling layer}. Let $r_{i}^\nu$, $\nu\in\{1, 2, 3\}$, be the spatial coordinates of the particle with position vector $r_i\in\mathbb{T}^3$. To simplify notation, we will also use $r_{i}^\nu$ to denote the inputs to layer $k$ (that is, the particles transformed by $M_{k-1}\circ\cdots\circ M_1$), and drop the dependence on $k$. Let $\mathbb{I}_k$ be a subset of the indices $\{1, 2, 3\}$ associated with layer $k$. Then, $M_k$ is defined as follows:
\begin{equation}\label{eq:coupling_layer}
    M_k\br{r_i^\nu} = \begin{cases}
        r_i^\nu & \nu \in \mathbb{I}_k\\
        G\br{r_i^\nu; \psi_i^{\nu}} & \nu\not\in \mathbb{I}_k,
    \end{cases}
\end{equation}
where
\begin{equation}\label{eq:conditioner}
    \psi_i^{\nu} = C_i^\nu\br{r_1^{\mathbb{I}_k}, \ldots, r_N^{\mathbb{I}_k}; \theta_k}.
\end{equation}
By $r_i^{\mathbb{I}_k}$ we refer to the set of coordinates of $r_i$ that are indexed by $\mathbb{I}_k$, and $C_i^\nu$ is the output of $C$ for coordinate $\nu$ of particle $i$. 
The functions $G$ and $C$ are implemented by neural networks. The parameters of $G$ associated with  coordinate $\nu$ of particle $i$ are denoted by $\psi_i^\nu$ and computed by $C$;
we have again dropped the dependence of $\psi_i^{\nu}$ on $k$ to simplify notation.
The parameters of $C$ are the learnable parameters of the layer, $\theta_k$.
An illustration of the coupling layer defined by Eqs.~(\ref{eq:coupling_layer}--\ref{eq:conditioner}) is shown in Fig.~\ref{fig:coupling_layer}.

In simple terms, $M_k$ works as follows. We partition the coordinates into two sets; the coordinates indexed by $\mathbb{I}_k$ are left invariant, whereas the remaining coordinates are transformed element-wise. Each transformed coordinate undergoes a different transformation depending on the value of $\psi_i^\nu$, which itself is a function of all the coordinates that remain invariant. To ensure that each coordinate gets the opportunity to be transformed as a function of every other coordinate, each layer $M_k$ uses a different partition $\mathbb{I}_k$, and we cycle through the partitions $\{1\}$, $\{2\}$, $\{3\}$, $\{1,2\}$, $\{2,3\}$ and $\{1,3\}$ across layers.

For the mapping $M$ to be bijective, a sufficient condition is that $G(\cdot; \psi): [-L,L]\rightarrow[-L,L]$ be strictly increasing for any setting of $\psi$. In that case, the inverse $M_k^{-1}$ is obtained by simply replacing $G$ with $G^{-1}$ in Eq.~(\ref{eq:coupling_layer}), and the Jacobian determinants can be computed efficiently as follows:
\begin{equation}\label{eq:coupling_layer_logdet}
    \log\abs{J_{M_k}} = \sum_{i=1}^N\sum_{\nu\not\in\mathbb{I}_k}\log \frac{\partial G}{\partial r_i^\nu}(r_i^\nu; \psi_i^\nu).
\end{equation}
Finally, the inverse and Jacobian determinant of the composite mapping $M$ can be computed by
\begin{align}
    M^{-1}& = M_1^{-1}\circ\ldots\circ M_K^{-1},\\
    \log\abs{J_{M}} &= \sum_{k=1}^K \log \abs{J_{M_k}}.
\end{align}

To ensure that the transformed distributions $\rho_{A'}$ and $\rho_{B'}$ obey the required boundary conditions, the implementation of $G$ must reflect the fact that $r_i^\nu=-L$ and $r_i^\nu=L$ are identified as the same point. For this to be the case, a sufficient set of conditions is the following:
\begin{align}
    G(\pm L; \psi) &= \pm L, \label{eq:circular_cond1}\\
    \frac{\partial G}{\partial r_i^\nu}(-L; \psi) &= \frac{\partial G}{\partial r_i^\nu}(L; \psi) > 0,\label{eq:circular_cond2}
\end{align}
for any setting of the parameters $\psi$. To satisfy the above conditions, we implement $G$ using \textit{circular splines}, which were recently proposed by Rezende et al.~\cite{rezende2020normalizing} and are based on the rational-quadratic spline flows of Durkan et al.~\cite{durkan2019neural}. Our implementation of the circular splines is detailed in Appendix~\ref{app:model}.

In addition to the above, we also need to make sure that the parameters $\psi_i^\nu$ in Eq.~(\ref{eq:conditioner}) are periodic functions of the invariant coordinates $r_1^{\mathbb{I}_k}, \ldots, r_N^{\mathbb{I}_k}$. This can be easily achieved by the following feature transformation:
\begin{equation}\label{eq:conditioner_input_transformation}
    r_i^{\mathbb{I}_k} \mapsto \br{\cos\br{\frac{\pi}{L} r_i^{\mathbb{I}_k}}, \sin\br{\frac{\pi}{L} r_i^{\mathbb{I}_k}}},
\end{equation}
where $\cos$ and $\sin$ act element-wise. The above feature transformation is injective, so no information about $r_i^{\mathbb{I}_k}$ is lost. We apply this feature transformation to each particle $i$ at the input layer of network $C$.

Finally, to ensure that the transformed distributions $\rho_{A'}$ and $\rho_{B'}$ are invariant to particle permutations, it is necessary that the Jacobian determinant $J_{M}$ also be invariant to particle permutations. In our architecture, this can be achieved by taking $C$ to be \textit{equivariant} to particle permutations. Specifically, let $\sigma$ be a permutation of the set $\{1, \ldots, N\}$. We say that $C$ is equivariant with respect to $\sigma$ if
\begin{equation}\label{eq:permutation_equivariance_conditioner}
    \psi_{\sigma(i)}^{\nu} = C_i^\nu\br{r_{\sigma(1)}^{\mathbb{I}_k}, \ldots, r_{\sigma(N)}^{\mathbb{I}_k}; \theta_k},
\end{equation}
that is, if permuting the particles has the effect of permuting the parameter outputs $\br{\psi_1^\nu, \ldots, \psi_N^\nu}$ in exactly the same way. From Eq.~(\ref{eq:coupling_layer_logdet}), we can easily see that the above property implies that $\sigma$ leaves $J_{M_k}$, and hence $J_M$, invariant, because we sum over all particles and the sum is permutation invariant. Previous studies have made similar observations~\cite{bender2020exchangeable, kohler2019}. Our implementation of $C$ is based on the architecture proposed by Vaswani et al.~\cite{vaswani2017transformer}, often referred to as the \textit{transformer}, which we use in a permutation-equivariant configuration. The implementation details of our transformer architecture are in Appendix~\ref{app:model}.

\section{\label{sec:results}Results}
In this section, we evaluate the performance of our method for the solvation system illustrated in Fig.~\ref{fig:system}. We focus on the bidirectional BAR and LBAR estimators in the main text, due to their advantages over unidirectional approaches as discussed in Sec.~\ref{sec:theory}\@. We refer to Appendix~\ref{app:results} for a discussion of the unidirectional counterparts.
\begin{figure}[htbp]
    \centering
    \includegraphics[width=.98\columnwidth]{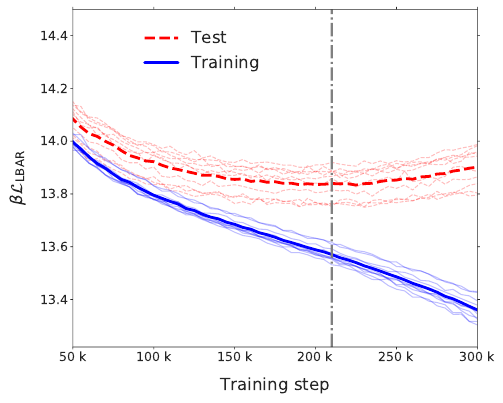}
    \caption{\textbf{Loss profile for bidirectional training.} The bidirectional loss [Eq.~(\ref{eq:loss_lbar})] is evaluated for training (solid lines) and test (dashed lines) datasets. Thin lines correspond to the individual runs, each trained on an independent dataset, and thick lines correspond to their respective averages. The loss keeps decreasing for the training set but exhibits a minimum for the test set at around \num{2.1e5}~steps (vertical line). This is roughly the point where the model starts to overfit to the training data.}
    \label{fig:loss}
\end{figure}

To capture statistical variation, our training and analysis procedure was performed $10$ times, each using independent training and evaluation datasets. Our loss profiles and free energy estimates below report averages as well as statistical variation across these runs.

We first report training results in Fig.~\ref{fig:loss}, where the full-batch loss is plotted as a function of the number of training steps. We observe a pattern commonly encountered in ML: after an initial decrease of both training and test loss, the latter develops a minimum. At around the minimum, the model stops generalizing and starts to overfit to the training data. We therefore employ a technique called \textit{early stopping}~\cite{caruana2001earlystopping} and use the model parameters corresponding to the minimum test loss for all further evaluations. It is worth emphasizing, however, that the precise location of the minimum does not have an appreciable effect on the quality of the free energy estimates reported for the bidirectional estimator (results not presented). We also note the small variation among the independent runs, suggesting that there is no significant dependence of the performance on a particular dataset.
\begin{figure*}[ht!]
    \centering
    \includegraphics[width=.49\linewidth]{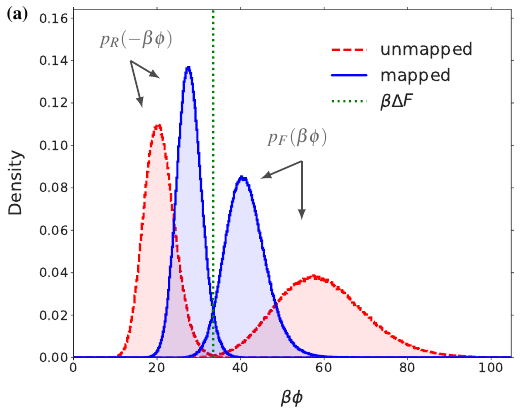}
    \hfill
    \includegraphics[width=.49\linewidth]{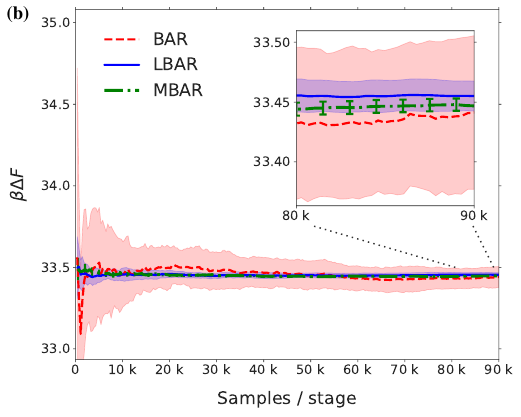}
    \caption{\textbf{Enhanced convergence of the learned LBAR estimator.} (a) Normalized histograms of forward and reverse work ($\beta \phi$) values with (blue, solid line) and without (red, dashed lines) the mapping. The vertical line indicates the ground-truth free energy estimate computed by MBAR\@. (b) Running averages of the $\Delta F$ estimate as a function of the number of evaluated samples per stage for the BAR (red, dashed lines), LBAR (blue, solid lines) and MBAR (green, dash-dotted lines) estimators. We note that the $x$-axis shows samples per stage; the MBAR estimate was computed using $15$ stages (as opposed to $2$ for BAR and LBAR) and thus in total it uses $7.5$ times more samples compared to the other two methods. The lines and shaded regions of BAR and LBAR estimates correspond to average and one standard deviation over the $10$ independent runs. The vertical green bars report statistical error of the MBAR estimator~\cite{Shirts2008}.}
    \label{fig:df}
\end{figure*}

Next, we probe the overlap resulting from our learned mapping. In Fig.~\ref{fig:df}a we plot the distributions of learned work values for the forward and reverse directions compared against their unmapped counterparts for which the mapping is the identity. These distributions are typically unimodal~\cite{cchipot07:molsim} and satisfy the inequalities ${-\mathbb E}_B[\Phi_R] \leq \Delta F \leq {\mathbb E}_{A}[\Phi_F]$
for any invertible mapping. The free energy difference $\Delta F$ corresponds precisely to the point of intersection of the forward and reverse distributions. Both of these facts are a consequence of the fluctuation theorem [Eq.~(\ref{eq:ft})]. Intuitively, it is clear then that an effective mapping should increase the overlap between the distributions to facilitate locating the intersection. This is precisely the enhancement we observe for the mapped work values. The two modes are shifted towards each other and share a significantly larger overlap than the unmapped distributions. We also see that the mapping strongly reduces the variance of the distributions. In fact, with a perfect mapping we would expect the forward and reverse distributions to collapse onto a single delta distribution located at $\Delta F$.

We now turn to the statistical convergence of the free energy estimates. 
In Fig.~\ref{fig:df}b we plot a running average of the estimate as a function of the number of evaluated samples per stage. The solid lines report averages of the estimate over the independent runs, and shaded regions represent one standard deviation of the runs.
We first validate the correctness of our method against a converged MBAR estimator. Here MBAR
 employed $15$ stages and thus $7.5$ times more samples in total. From the figure we see that the variation of our estimate overlaps nicely with the MBAR error estimates.
We next compare the efficiency of our method against the baseline BAR estimator, where we see in Fig.~\ref{fig:df}b clear variance reduction of LBAR across a wide range of evaluated sample sizes. The full-batch LBAR standard deviation we observe is approximately $19\%$ of that reported by BAR\@.
Moreover, training and evaluation of the model occurred on the same dataset, demonstrating that an effective mapping can be learned in a \textit{data-efficient} manner. This is an important practical consideration but not at all obvious a priori. We could, in principle, even combine samples from the training and test datasets for estimation of $\Delta F$ but have used the test set only to detect overfitting. 

\section{\label{sec:discussion}Discussion}
In this work, we turn TFEP into a machine learning problem by combining it with state-of-the-art ML techniques. TFEP previously required hand-crafting a tractable mapping on configuration-space, a significant challenge for many realistic systems. We proposed to represent the mapping by a suitable neural network, and identified training objectives for unidirectional and bidirectional cases. We then tested their performance on a prototype solvation system -- the growth of a soft sphere in a fluid of solvent particles in periodic boundary conditions. While this system is relatively simplistic from a physical standpoint, it poses a significant challenge for ML models due to the system's underlying permutational symmetry and periodic boundary conditions. Our experimental results indicate that both LFEP and LBAR estimators can lead to a significant variance reduction compared to their respective baselines and therefore clearly highlight the potential of this approach. Interesting directions for future work include a systematic analysis of the error of the learned estimators and a detailed comparison with MBAR. We believe it is possible that optimal estimation strategies on complex systems will contain a combination of staging and mapping.

Improving TFEP via learned mappings relates to the general idea of improving importance sampling by learning the proposal distribution, which has been explored substantially in machine learning and statistics. For instance, recent works in machine learning have proposed training a flexible deep-learning model of the proposal distribution to improve importance sampling~\cite{Mueller2018} or more sophisticated variants such as bridge sampling~\cite{papamakarios2015distilling} and sequential Monte Carlo~\cite{gu2015nasmc,paige2016inferencenets,le2017inferencecompilation}. In turn, these approaches can be traced back to methods for adaptive importance sampling~\cite{cappe2008adaptive} and adaptive sequential Monte Carlo~\cite{cornebise2008adaptive} in statistics. One recent instance of these approaches that relates closely to our work is \textit{Neural Importance Sampling}~\cite{Mueller2018}, which uses expressive normalizing flows to learn good proposal distributions for importance sampling. Many of the above works have noted that the choice of loss function is important, with the forward KL divergence and chi-squared divergence being standard choices. These observations are in line with our observations of the differences between the unidirectional and bidirectional training losses.

We note that our learned free energy estimators and equivariant, periodic model architecture can be combined with the work on \textit{Boltzmann Generators}~\cite{Noe19:Science}, which uses flow-based models in combination with statistical re-weighting to sample from desired Boltzmann distributions. In particular, it would be interesting to see how our targeted estimators compare to the ones considered in Ref.~\onlinecite{Noe19:Science}. Furthermore, we note that neural network based free energy estimation is an active field of research. For example, two recent studies have independently proposed targeted unidirectional~\cite{Nicoli2020} and bidirectional~\cite{Ding2020} estimators similar to the ones suggested here. Both studies employ autoregressive networks to compute free energy estimates of a lattice spin model with discrete states. In addition, Ref.~\onlinecite{Ding2020} also estimates free energies of a small protein in the gas phase (no periodicity) using normalizing flows. As noted in that study, however, their model lacks permutation equivariance which is likely a drawback when applying it to the type of solvation system we consider here. We also believe that permutation equivariant, periodicity-respecting networks, such as the one considered here, will be key to scaling up the approach to system sizes commonly used in atomistic simulations. 

Finally, our results demonstrate that we can estimate free energy differences between two states directly and data efficiently, i.e.\ using fewer MD samples for training and evaluation than the base estimator would require to converge. Other studies, for example Refs.~\onlinecite{Noe19:Science, Ding2020}, follow a different strategy and estimate free energy differences by learning two separate mappings that share a common reference state. We believe that our direct approach may be preferable in cases where one state is a small perturbation of the other. It will be interesting to see how these different approaches compare with respect to data efficiency and variance reduction, and under which circumstances one is preferable to the other.

\begin{acknowledgments}
    We would like to thank our colleagues Shakir Mohamed, Michael Figurnov, Ulrich Paquet, James Kirkpatrick, Daan Wierstra, Craig Donner, Alex Vitvitskyi, John Jumper, Amber Nicklin-Clark, Steph Hughes-Fitt, Alex Goldin and Guy Scully for their help and for stimulating discussions.
\end{acknowledgments}

\appendix
\section{\label{app:tfepder} Alternate derivation of TFEP and interpretation}

In this section we derive and interpret the unidirectional TFEP estimator as a
multi-staged FEP estimator. This interpretation allows us to reason about TFEP
using intuition from FEP\@.

Given explicit densities for $A$, $A^\prime$ and $B$, we can formally decompose
$\Delta F$ into a sum over two terms,
\begin{equation}
  \Delta F = \Delta F_{AA^\prime} + \Delta F_{A^\prime B} , \label{eq:dFdecomp}
\end{equation}
which can each be computed separately using the FEP estimator Eq.~(\ref{eq:fep}). We next define
the energy of $A^\prime$ as
\begin{align}
  U_{A^\prime} (\mathbf{x}) & =  
   U_A \br{M^{- 1} (\mathbf{x})} + \beta^{- 1} \log \abs{J_M (\mathbf{x})} 
  \nonumber\\
  & = - \beta^{- 1} \log \left[ \rho_{A^\prime} (\mathbf{x}) Z_A \right] \label{eq:UAp}, 
\end{align}
such that $\rho_{A^\prime} \propto e^{- \beta U_{A^\prime}}$, and where we have
used Eq.~(\ref{eq:change_of_variableA}) in going from the first to second line. Conveniently, one of these stages comes for free, as $\Delta
F_{AA^\prime} = 0$:
\begin{align}
  Z_{A^\prime} & = \int \mathrm{d} \mathbf{x}\ e^{- \beta U_{A^\prime} (\mathbf{x})} \nonumber\\
& = Z_A \int \mathrm{d} \mathbf{x}\ \rho_{A^\prime} (\mathbf{x}) \nonumber\\
  & = Z_A . \label{eq:AAprimeequiv}
\end{align}
In going from the first to second line we have used Eq.~(\ref{eq:UAp}). Combining Eqs.~(\ref{eq:dFdecomp}--\ref{eq:AAprimeequiv}), as well as the FEP estimator Eq.~(\ref{eq:fep}), we arrive at our final result:
\begin{align}
  e^{- \beta \Delta F} & =  e^{- \beta \Delta F_{A^{\prime} B}}\nonumber\\
  & = \mathbb{E}_{A^{\prime}} \left[e^{- \beta \Delta U^{\prime}}\right],
  \label{eq:tfep_reformulation}
\end{align}
where
\begin{align}
  \Delta U^{\prime} (\mathbf{x}) &= U_B (\mathbf{x}) - U_{A^{\prime}} (\mathbf{x})\nonumber\\
  &= U_B (\mathbf{x}) - U_A \br{M^{- 1} (\mathbf{x})} - \beta^{- 1} \log \abs{J_{M}
  (\mathbf{x})} \label{eq:dUprime}
\end{align}
is the energy difference between $A^{\prime}$ and $B$. Although Eq.~(\ref{eq:tfep_reformulation}) is an
explicit estimate between $A^{\prime}$ and $B$, it is just a reformulation of the TFEP estimator 
[Eq.~(\ref{eq:tfep}) above] as can be seen by the equivalence of $\Delta U^\prime$ and $\Phi_F$ [compare Eqs.~(\ref{eq:phi_f}) and (\ref{eq:dUprime})]. This interpretation of TFEP allows us to apply the intuition on convergence we have built for FEP\@. Specifically, we can accelerate convergence if $A^{\prime}$ shares large overlap with $B$.

\section{\label{app:system}System}
To generate the training data, we performed MD simulations of the system illustrated in Fig.~\ref{fig:system} using the simulation package LAMMPS~\cite{Plimpton1995}. The system is similar to the one studied in Ref.~\onlinecite{jarzynski02:pre} but we replaced hard solute-solvent interactions by a Weeks--Chandler--Andersen~\cite{Weeks1971} (WCA) potential. Below, we represent our energy, length, and mass units in terms of the LJ well depth $\epsilon$, the LJ diameter $\sigma$, and the solvent particle mass $m$~\cite{Frenkel2002}. From this our unit of time is defined as $\tau = \sigma \sqrt{m / \epsilon}$. Quantities expressed in these reduced units are denoted with an asterisk. 
We used a cubic simulation box with edge length $2L^{*}=6.29$, employed cutoff radii of $L^{*}$ and $\sqrt[\leftroot{-2}\uproot{2}6]{2} R_\alpha$ for LJ and WCA interactions, where $\alpha \in \{A, B\}$ labels the state. The solute radii were taken to be $R_A^{*} = 2.5974$ and $R_B^{*} = 2.8444$. Both LJ and WCA potentials shared the same value for $\epsilon$, and we set $\sigma_{\text{WCA}} = R_{\alpha}$.

To simulate a specific state, we first assigned random positions to all solvent particles. The solute was placed at the origin of the box and kept stationary throughout the simulation. After performing energy minimization, the system was equilibrated in the canonical ensemble for a period of $5 \times 10^4 \tau$. The equations of motion were integrated using the velocity Verlet algorithm~\cite{Swope1982} with a timestep $0.002 \tau$. We employed a Langevin thermostat with a relaxation time of $0.5\tau$ to keep the system at a temperature of $T^{*} = 3.2155$. To prevent drift of the center of mass motion, we set the total random force to zero. During the $5 \times 10^5 \tau$ long production run, configurations were sampled every $10$ reduced time units, yielding a total of \num{5e4} samples. For each state, $20$ such simulations were generated starting from random initializations and random number seeds. The resulting \num{1e6} samples were then partitioned as described in the main text. 

We followed the same protocol to generate samples for MBAR\@. In addition to the two states corresponding to $R_A$ and $R_B$, we considered $13$ intermediate states with a constant radial increment $\Delta R_{i, i+1}$. Using two different random seeds, we obtained \num{1e5} samples for each state and evaluated all $15$ energy functions on each sample. MBAR estimates of $\Delta F$ were then computed from this combined energy matrix using the package pymbar~\cite{Shirts2008}.

\section{\label{app:model}Model implementation details}

We implement $G$ using \textit{circular splines}~\cite{rezende2020normalizing} so that $G$ satisfies the boundary conditions in Eqs.~(\ref{eq:circular_cond1}--\ref{eq:circular_cond2}).
Briefly, $G$ is a piecewise function that consists of $S$ segments. The parameters $\psi$ are a set of $S+1$ triplets $(x_s, y_s, d_s)$, where $[x_{s-1}, x_s]$ defines the domain of segment $s$, $[y_{s-1}, y_s]$ defines its image, and $d_{s-1}, d_s$  define its slopes at the endpoints (all slopes are required to be positive). Each segment is a strictly increasing rational-quadratic function, constructed using the interpolation method of Gregory and Delbourgo~\cite{gregory1982rationalquadratic}. The conditions in Eqs.~(\ref{eq:circular_cond1}--\ref{eq:circular_cond2}) are satisfied by setting $x_0 = y_0 = -L$, $x_S = y_S = L$ and $d_0 = d_S > 0$. We can increase the flexibility of $G$ by increasing the number of segments $S$. With a large enough $S$, circular splines can approximate arbitrarily well any strictly increasing function from $[-L, L]$ to itself that satisfies Eqs.~(\ref{eq:circular_cond1}--\ref{eq:circular_cond2}).

We implement $C$ using the \textit{transformer architecture} of Vaswani et al.~\cite{vaswani2017transformer} so that $C$ satisfies the permutation-equivariance condition in Eq.~(\ref{eq:permutation_equivariance_conditioner}). Briefly, the network architecture is as follows. Each input $r_i^{\mathbb{I}_k}$ undergoes the feature transformation in Eq.~(\ref{eq:conditioner_input_transformation}), and is then mapped to a fixed-sized vector $h_i$ via an affine transformation identically for each $i$. Then, $(h_1, \ldots, h_N)$ is processed by a sequence of \textit{transformer blocks}, each of which is composed of two \textit{residual layers}~\cite{he2016resnet}. The first residual layer is
\begin{equation}
    h_i \leftarrow h_i + \mathrm{MHA}_i\br{h_1, \ldots, h_N},
\end{equation}
where $\mathrm{MHA}$ is a \textit{multi-headed self-attention layer} as described by Vaswani et al.~\cite{vaswani2017transformer}, and $\mathrm{MHA}_i$ is its $i$-th output. The second residual layer is
\begin{equation}
    h_i \leftarrow h_i + \mathrm{MLP}\br{h_i},
\end{equation}
where $\mathrm{MLP}$ is a standard \textit{multi-layer perceptron}~\cite{goodfellow2016deeplearningbook} that is applied identically for each $i$. After repeating the transformer block a specified number of times (each time with different learnable parameters), each vector $h_i$ is mapped to the output $\br{\psi_i^\nu}_{\nu\not\in \mathbb{I}_k}$ via an affine transformation identically for each $i$.
To help with generalization and stability, we applied \textit{layer normalization}~\cite{ba2016layernorm} to the inputs of $\mathrm{MHA}$ and $\mathrm{MLP}$, as well as the output\@. Because $\mathrm{MHA}$ is permutation-equivariant and every $\mathrm{MLP}$ and affine transformation is applied identically for each $i$, it follows that $C$ is permutation-equivariant, and therefore the transformed distribution is permutation-invariant as desired.

All models were trained using the Adam optimizer~\cite{Kingma2014}. A summary of the hyperparameters is provided in Table~\ref{tab:hyper}. 

\begin{table}[ht]
\caption{\label{tab:hyper}Model and training hyperparameters.}
\begin{ruledtabular}
\begin{tabular}{ll} 
\textbf{Transformer} & \\ \hline
Number of blocks & 4\\
Number of heads & 4 \\ 
Value dimension & 128 \\
Key dimension & 128 \\
Embedding dimension & 128 \\ 
\\
\textbf{Circular spline flow} & \\ \hline
Number of segments ($S$) & 32\\ 
\\
\textbf{Adam optimizer} & \\ \hline
Batch size & 256 \\
Learning rate & $10^{-4}$ \\
$\beta_1$ & 0.9 \\ 
$\beta_2$ & 0.999\\
\end{tabular}
\end{ruledtabular}
\end{table}

\begin{figure*}[htbp]
    \centering
    \includegraphics[width=.49\linewidth]{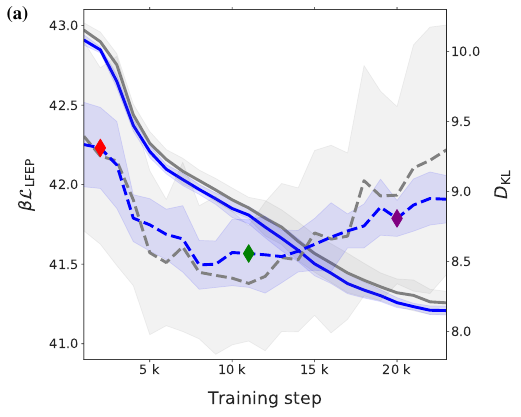}
    \hfill
    \includegraphics[width=.49\linewidth]{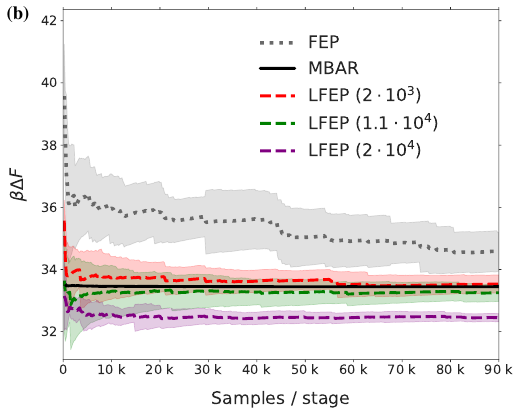}
    \caption{\textbf{Enhanced convergence of the learned LFEP estimator.} (a) Estimates of the loss $\mathcal L_{\text{LFEP}}$ (solid lines) and $D_{\text{KL}}$ (dashed lines), averaged across $10$ different datasets, as a function of training progress for test (grey) and training (blue) datasets. Shaded regions correspond to one standard deviation estimated across the independent runs. Colored diamonds highlight selected training steps for evaluation of $\Delta F$. (b)
    Running averages of the $\Delta F$ estimate as a function of the number of evaluated samples per stage for the FEP (dotted line), MBAR (solid line) and LFEP (dashed lines) estimators. For the latter, the numbers within parentheses correspond to the training steps at which we evaluated the mapping. The lines and shaded regions of FEP and LFEP estimates correspond to average and one standard deviation over the runs.}
    \label{fig:dflfep}
\end{figure*}

\section{\label{app:results}Results for LFEP}
In this section, we discuss the free energy estimates obtained with LFEP, using unidirectional training with the $\mathcal L_{\text{LFEP}}$ loss [Eq.~(\ref{eq:loss_lfep})] as a proxy for $D_\text{KL}$ [Eq.~(\ref{eq:lossder})]. To estimate the KL, we replaced  $\Delta F$ in Eq.~(\ref{eq:lossder}) with the LFEP estimate of that quantity. Figure~\ref{fig:dflfep}a shows the evolution of both quantities during training. The results feature an interesting behaviour in the initial training regime. While test and training loss are still decreasing, the KL already exhibits a pronounced minimum due to a drift of the $\Delta F$ estimate towards lower values. This is further illustrated in Fig.~\ref{fig:dflfep}b, which compares the convergence of $\Delta F$ for three different mappings corresponding to specific training steps (colored symbols in Fig.~\ref{fig:dflfep}a). We see that the variance is reduced significantly in all three cases. However, only the first mapping (training step \num{2e3}) agrees with the MBAR baseline, 
while we can already observe a small bias for the second mapping (training step \num{1.1e4}) 
that becomes even more pronounced as training progresses (training step \num{2e4}). We note that this behaviour is consistent with other experiments that we performed in the unidirectional setting (data not shown).

This is problematic in that the minimum of the KL would be a natural point to stop training but does not yield the lowest bias. This is also in contrast to our findings for the bidirectional case, where the quality of the mapping was best in the vicinity of the minimum in the test loss. 
One possible explanation for these observations is the well-known zero-forcing property~\cite{minka2005divergence} of $\mathcal L_{\text{LFEP}}$. That is, $\mathcal L_{\text{LFEP}}$ does not encourage the transformed distribution to be mass-covering, which is known to negatively impact the performance of importance-sampling estimators~\cite{neal2005bridgesampling}.

\bibliography{document}

\end{document}